\documentclass[preliminary]{eptcs}

\makeatletter
\@ifundefined{lhs2tex.lhs2tex.sty.read}%
  {\@namedef{lhs2tex.lhs2tex.sty.read}{}%
   \newcommand\SkipToFmtEnd{}%
   \newcommand\EndFmtInput{}%
   \long\def\SkipToFmtEnd#1\EndFmtInput{}%
  }\SkipToFmtEnd

\newcommand\ReadOnlyOnce[1]{\@ifundefined{#1}{\@namedef{#1}{}}\SkipToFmtEnd}
\usepackage{amstext}
\usepackage{amssymb}
\usepackage{stmaryrd}
\DeclareFontFamily{OT1}{cmtex}{}
\DeclareFontShape{OT1}{cmtex}{m}{n}
  {<5><6><7><8>cmtex8
   <9>cmtex9
   <10><10.95><12><14.4><17.28><20.74><24.88>cmtex10}{}
\DeclareFontShape{OT1}{cmtex}{m}{it}
  {<-> ssub * cmtt/m/it}{}

\DeclareFontShape{OT1}{cmtt}{bx}{n}
  {<5><6><7><8>cmtt8
   <9>cmbtt9
   <10><10.95><12><14.4><17.28><20.74><24.88>cmbtt10}{}
\DeclareFontShape{OT1}{cmtex}{bx}{n}
  {<-> ssub * cmtt/bx/n}{}

\newcommand{\Conid}[1]{\mathit{#1}}
\newcommand{\Varid}[1]{\mathit{#1}}
\newcommand{\anonymous}{\kern0.06em \vbox{\hrule\@width.5em}}
\newcommand{\plus}{\mathbin{+\!\!\!+}}

\usepackage{polytable}

\@ifundefined{mathindent}%
  {\newdimen\mathindent\mathindent\leftmargini}%
  {}%

\def\resethooks{%
  \global\let\SaveRestoreHook\empty
  \global\let\ColumnHook\empty}
\newcommand*{\savecolumns}[1][default]%
  {\g@addto@macro\SaveRestoreHook{\savecolumns[#1]}}
\newcommand*{\restorecolumns}[1][default]%
  {\g@addto@macro\SaveRestoreHook{\restorecolumns[#1]}}
\newcommand*{\aligncolumn}[2]%
  {\g@addto@macro\ColumnHook{\column{#1}{#2}}}

\resethooks

\newcommand{\onelinecommentchars}{\quad-{}- }
\newcommand{\commentbeginchars}{\enskip\{-}
\newcommand{\commentendchars}{-\}\enskip}

\newcommand{\visiblecomments}{%
  \let\onelinecomment=\onelinecommentchars
  \let\commentbegin=\commentbeginchars
  \let\commentend=\commentendchars}

\newcommand{\invisiblecomments}{%
  \let\onelinecomment=\empty
  \let\commentbegin=\empty
  \let\commentend=\empty}

\visiblecomments

\newlength{\blanklineskip}
\newcommand{\hsindent}[1]{\quad}
\let\hspre\empty
\let\hspost\empty

\EndFmtInput
\makeatother

\providecommand{\event}{TFPIE 2014} 
\usepackage{breakurl}               
\usepackage{color}
\usepackage{graphicx}
\usepackage{verbatim}

\newcommand{\askelle}{{\sc Ask-Elle}}
\newcommand{\ideas}{{\sc Ideas}}
\newcommand{\ignore}[1]{}
\newenvironment{spaced}{\addtolength{\parskip}{0.7\baselineskip}}{}
\title{Evaluating Haskell expressions in a tutoring environment}

\author{
Tim Olmer
\quad\quad Bastiaan Heeren 
\institute{Open University of the Netherlands\\ Faculty of Management, Science and Technology\\ Heerlen, The Netherlands}
\email{tim.olmer@gmail.com \quad\quad bhr@ou.nl}
\and Johan Jeuring
\institute{Universiteit Utrecht\\ Department of Information and Computing Sciences\\ Utrecht, The Netherlands}
\email{J.T.Jeuring@uu.nl}
}

\def\titlerunning{Evaluating Haskell expressions in a tutoring environment}
\def\authorrunning{T. Olmer, B. Heeren \& J. Jeuring}

\begin{document}
\maketitle

\begin{abstract}
A number of introductory textbooks for Haskell use calculations right from the
start to give the reader insight into the evaluation of expressions and the
behavior of functional programs. Many programming concepts that are important in
the functional programming paradigm, such as recursion, higher-order functions,
pattern-matching, and lazy evaluation, can be partially explained by showing a
stepwise computation. A student gets a better understanding of these concepts
if she performs these evaluation steps herself. Tool support for experimenting
with the evaluation of Haskell expressions is currently lacking. In this paper
we present a prototype implementation of a stepwise evaluator for Haskell
expressions that supports multiple evaluation strategies, specifically targeted
at education. Besides performing evaluation steps the tool also diagnoses
steps that are submitted by a student, and provides feedback. Instructors 
can add or change function definitions without knowledge of the tool's 
internal implementation. We discuss some preliminary results of a small
survey about the tool.
\end{abstract}

\section{Introduction}

Many textbooks that introduce the functional programming language Haskell begin 
with showing calculations that illustrate how expressions are evaluated, 
emphasizing the strong correspondence with mathematical expressions and the
property of referential transparency of the language. For instance, Hutton 
presents calculations for the expressions \ensuremath{\Varid{double}\;\mathrm{3}} and \ensuremath{\Varid{double}\;(\Varid{double}\;\mathrm{2})} on
page~1 of his textbook~\cite{Hutton}, and Bird and Wadler show the steps of 
reducing \ensuremath{\Varid{square}\;(\mathrm{3}\mathbin{+}\mathrm{4})} on page~5 of their book~\cite{Bird}. Similar 
calculations can be found in the first chapter of Thompson's The Craft of 
Functional Programming~\cite{Thompson} and Hudak's The Haskell School of 
Expression. Textbooks that do not show these calculations~\cite{RWH, LearnYouAHaskell}
compensate for this by giving lots of examples with an interpreter.

\begin{figure}[t]
\hfill\includegraphics[width=15cm]{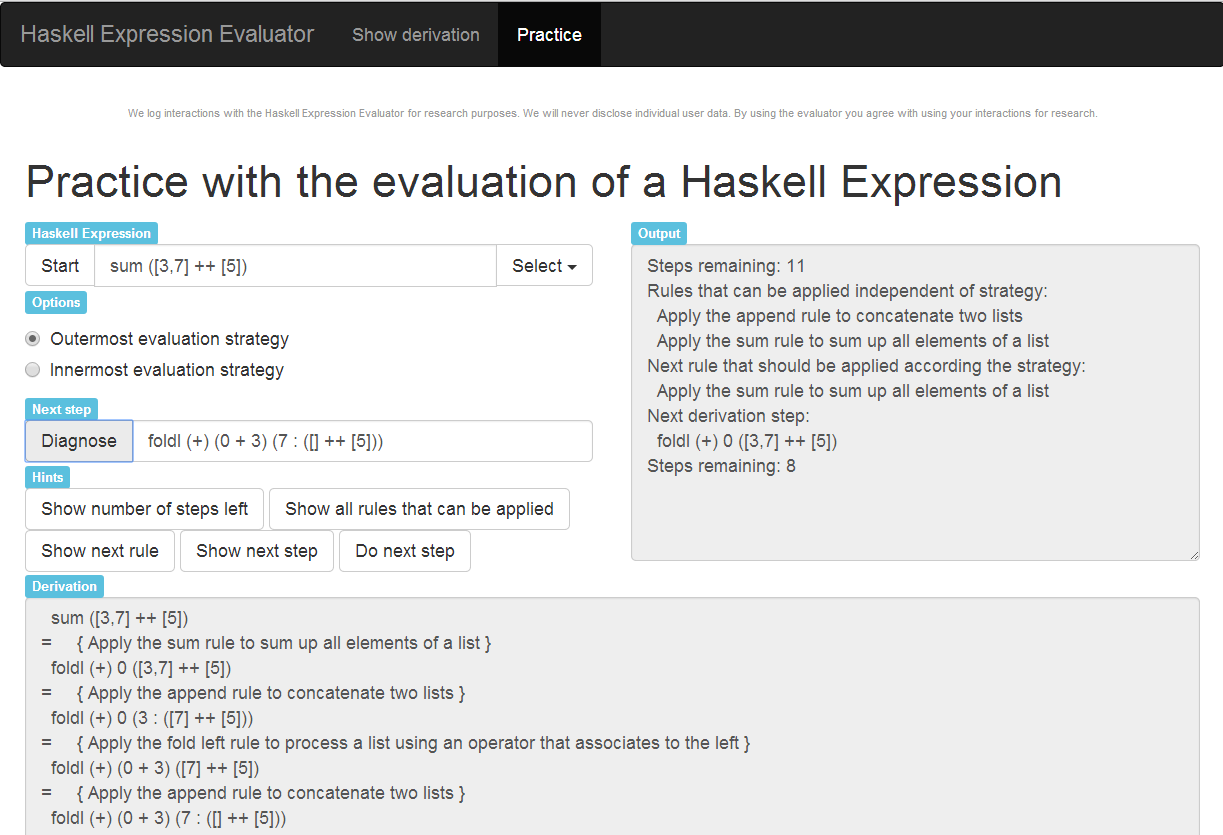}\hfill\mbox{}
\caption{Prototype screen for student interaction}
\label{fig:prototype}
\end{figure}

Stepwise evaluating an expression on a piece of paper can give a student a 
feeling for what a program does~\cite{chakravarty2004risks}.
However, there is no simple way to view intermediate evaluation steps for a
Haskell expression. In this paper we present a prototype implementation of the 
Haskell Expression Evaluator (HEE) that can show evaluation steps, and lets 
students practice with evaluating expressions on their own by providing feedback
and suggestions (see Figure~\ref{fig:prototype}).\footnote{The prototype is available via \url{http://ideas.cs.uu.nl/HEE/}.}
The tool supports multiple evaluation strategies and can handle
multiple (alternative) definitions for functions from the prelude. It is 
relatively easy for instructors to change the granularity of the steps, or to 
customize the feedback messages that are reported by the tool. 

Showing calculations can be a useful approach to let a student better understand 
some of the central programming concepts behind the programming language Haskell,
such as pattern-matching, recursion, higher-order functions, and lazy evaluation.
This approach is also used in textbooks
on Haskell~\cite{Hutton, Bird}. For instance, Haskell's lazy evaluation strategy
is fundamentally different from other mainstream programming languages 
(such as C, C++, C\# and Java), and the tool can highlight these differences.
A novice functional programmer often faces difficulties in understanding the 
evaluation steps in a lazy language, and even more experienced programmers
find it hard to predict the space behavior of their programs~\cite{Bakewell:space}.
Another stumbling block is the very compact syntax that is used in Haskell, 
which makes it sometimes hard to get an operational view of a functional
program~\cite{sparud:redex}. More generally, students often do not clearly 
understand operator precedence and associativity and misinterpret expressions~\cite{Krishna,Kumar}.
Showing evaluation steps can partly alleviate these problems.

\paragraph{Contributions and scope.} 

This paper presents a tool that enables students to
practice with evaluation steps for Haskell expressions. A student can not only
inspect the evaluation steps of a program, but can also provide evaluation steps as
input, which can then be checked against various evaluation strategies.
Furthermore, the steps are presented at a level of abstraction typically
expected in an educational setting. Practicing with evaluation steps gives
a student insight into how certain programming concepts such as recursion,
higher-order functions, and pattern matching work. It also gives a student
insight into various evaluation strategies.

Our evaluation tool only supports integers, list notation, recursion, higher
order functions, and pattern matching. The target audience for the evaluator is
students taking an introductory course on functional programming. Another
limitation is that only small code fragments are considered, which is sufficient
for the intended audience. We assume that the expressions evaluated by the tool 
are well-typed and do not contain compile-time errors, although we could let the
tool check for errors before evaluating.

Related work mainly focuses on showing evaluation
steps~\cite{Launchbury93anatural, Rochel:2009:VL9:1929087.1929099}, and does not
offer the possibility to let a student enter evaluation steps herself, or
presents evaluation steps at a lower level of abstraction, such as the
lambda-calculus~\cite{Sestoft02demonstratinglambda,
Abadi96explicitsubstitutions}.

\paragraph{Roadmap.}
The rest of the paper is structured as follows. We start with an example that
illustrates different evaluation strategies in Section~\ref{section:example}.
Next, we define rewrite rules and rewrite strategies for a simple expression 
language in Section~\ref{section:rewriting}, which are used for stepwise
evaluating an expression and for calculating feedback. 
We also discuss how rewrite rules and rewrite strategies can be generated for
arbitrary function definitions.
Section~\ref{section:feedback}
discusses the prototype in more detail, shows how we present feedback,
and describes the results of the survey.
We conclude the paper with a discussion on related work (Section~\ref{section:relatedwork})
and present conclusions and future work (Section~\ref{section:conclusion}).

\section{An example}
\label{section:example}

We start by demonstrating some evaluation strategies that are supported by the tool.
We use the expression \ensuremath{\Varid{sum}\;([\mskip1.5mu \mathrm{3},\mathrm{7}\mskip1.5mu]\plus [\mskip1.5mu \mathrm{5}\mskip1.5mu])} as a running example in the rest of the 
paper. 
Because the tool is developed for education, we use list notation when showing
expressions and we make associativity explicit. The evaluation steps in our examples are based on the following
standard definitions for prelude functions:

\begin{spaced}\begingroup\par\noindent\advance\leftskip\mathindent\(
\begin{pboxed}\SaveRestoreHook
\column{B}{@{}>{\hspre}l<{\hspost}@{}}%
\column{9}{@{}>{\hspre}l<{\hspost}@{}}%
\column{10}{@{}>{\hspre}l<{\hspost}@{}}%
\column{13}{@{}>{\hspre}l<{\hspost}@{}}%
\column{21}{@{}>{\hspre}l<{\hspost}@{}}%
\column{E}{@{}>{\hspre}l<{\hspost}@{}}%
\>[B]{}\Varid{sum}\mathbin{::}[\mskip1.5mu \Conid{Int}\mskip1.5mu]\to \Conid{Int}{}\<[E]%
\\
\>[B]{}\Varid{sum}\mathrel{=}\Varid{foldl}\;(\mathbin{+})\;\mathrm{0}{}\<[E]%
\\[\blanklineskip]%
\>[B]{}\Varid{foldl}\mathbin{::}(\Varid{a}\to \Varid{b}\to \Varid{a})\to \Varid{a}\to [\mskip1.5mu \Varid{b}\mskip1.5mu]\to \Varid{a}{}\<[E]%
\\
\>[B]{}\Varid{foldl}\;\Varid{f}\;{}\<[10]%
\>[10]{}\Varid{v}\;{}\<[13]%
\>[13]{}[\mskip1.5mu \mskip1.5mu]{}\<[21]%
\>[21]{}\mathrel{=}\Varid{v}{}\<[E]%
\\
\>[B]{}\Varid{foldl}\;\Varid{f}\;{}\<[10]%
\>[10]{}\Varid{v}\;{}\<[13]%
\>[13]{}(\Varid{x}\mathbin{:}\Varid{xs}){}\<[21]%
\>[21]{}\mathrel{=}\Varid{foldl}\;\Varid{f}\;(\Varid{f}\;\Varid{v}\;\Varid{x})\;\Varid{xs}{}\<[E]%
\\[\blanklineskip]%
\>[B]{}(\plus )\mathbin{::}[\mskip1.5mu \Varid{a}\mskip1.5mu]\to [\mskip1.5mu \Varid{a}\mskip1.5mu]\to [\mskip1.5mu \Varid{a}\mskip1.5mu]{}\<[E]%
\\
\>[B]{}[\mskip1.5mu \mskip1.5mu]{}\<[9]%
\>[9]{}\plus \Varid{ys}\mathrel{=}\Varid{ys}{}\<[E]%
\\
\>[B]{}(\Varid{x}\mathbin{:}\Varid{xs}){}\<[9]%
\>[9]{}\plus \Varid{ys}\mathrel{=}\Varid{x}\mathbin{:}(\Varid{xs}\plus \Varid{ys}){}\<[E]%
\ColumnHook
\end{pboxed}
\)\par\noindent\endgroup\resethooks
\end{spaced}%

\begin{figure}[t]
\begin{displaymath}
\begin{minipage}{7.5cm}
\begingroup\par\noindent\advance\leftskip\mathindent\(
\begin{pboxed}\SaveRestoreHook
\column{B}{@{}>{\hspre}c<{\hspost}@{}}%
\column{BE}{@{}l@{}}%
\column{6}{@{}>{\hspre}l<{\hspost}@{}}%
\column{8}{@{}>{\hspre}l<{\hspost}@{}}%
\column{E}{@{}>{\hspre}l<{\hspost}@{}}%
\>[B]{}\hspace*{1cm}{}\<[BE]%
\>[6]{}\Varid{sum}\;([\mskip1.5mu \mathrm{3},\mathrm{7}\mskip1.5mu]\plus [\mskip1.5mu \mathrm{5}\mskip1.5mu]){}\<[E]%
\\
\>[B]{}\mathrel{=}{}\<[BE]%
\>[8]{}\hspace*{5mm}\{\ \textrm{definition }\textit{sum}\ \}{}\<[E]%
\\
\>[B]{}\hspace*{1cm}{}\<[BE]%
\>[6]{}\Varid{foldl}\;(\mathbin{+})\;\mathrm{0}\;([\mskip1.5mu \mathrm{3},\mathrm{7}\mskip1.5mu]\plus [\mskip1.5mu \mathrm{5}\mskip1.5mu]){}\<[E]%
\\
\>[B]{}\mathrel{=}{}\<[BE]%
\>[8]{}\hspace*{5mm}\{\ \textrm{definition $+\!+$}\ \}{}\<[E]%
\\
\>[B]{}\hspace*{1cm}{}\<[BE]%
\>[6]{}\Varid{foldl}\;(\mathbin{+})\;\mathrm{0}\;(\mathrm{3}\mathbin{:}([\mskip1.5mu \mathrm{7}\mskip1.5mu]\plus [\mskip1.5mu \mathrm{5}\mskip1.5mu])){}\<[E]%
\\
\>[B]{}\mathrel{=}{}\<[BE]%
\>[8]{}\hspace*{5mm}\{\ \textrm{definition \textit{foldl}}\ \}{}\<[E]%
\\
\>[B]{}\hspace*{1cm}{}\<[BE]%
\>[6]{}\Varid{foldl}\;(\mathbin{+})\;(\mathrm{0}\mathbin{+}\mathrm{3})\;([\mskip1.5mu \mathrm{7}\mskip1.5mu]\plus [\mskip1.5mu \mathrm{5}\mskip1.5mu]){}\<[E]%
\\
\>[B]{}\mathrel{=}{}\<[BE]%
\>[8]{}\hspace*{5mm}\{\ \textrm{definition $+\!+$}\ \}{}\<[E]%
\\
\>[B]{}\hspace*{1cm}{}\<[BE]%
\>[6]{}\Varid{foldl}\;(\mathbin{+})\;(\mathrm{0}\mathbin{+}\mathrm{3})\;(\mathrm{7}\mathbin{:}([\mskip1.5mu \mskip1.5mu]\plus [\mskip1.5mu \mathrm{5}\mskip1.5mu])){}\<[E]%
\\
\>[B]{}\mathrel{=}{}\<[BE]%
\>[8]{}\hspace*{5mm}\{\ \textrm{definition \textit{foldl}}\ \}{}\<[E]%
\\
\>[B]{}\hspace*{1cm}{}\<[BE]%
\>[6]{}\Varid{foldl}\;(\mathbin{+})\;((\mathrm{0}\mathbin{+}\mathrm{3})\mathbin{+}\mathrm{7})\;([\mskip1.5mu \mskip1.5mu]\plus [\mskip1.5mu \mathrm{5}\mskip1.5mu]){}\<[E]%
\\
\>[B]{}\mathrel{=}{}\<[BE]%
\>[8]{}\hspace*{5mm}\{\ \textrm{definition $+\!+$}\ \}{}\<[E]%
\\
\>[B]{}\hspace*{1cm}{}\<[BE]%
\>[6]{}\Varid{foldl}\;(\mathbin{+})\;((\mathrm{0}\mathbin{+}\mathrm{3})\mathbin{+}\mathrm{7})\;[\mskip1.5mu \mathrm{5}\mskip1.5mu]{}\<[E]%
\\
\>[B]{}\mathrel{=}{}\<[BE]%
\>[8]{}\hspace*{5mm}\{\ \textrm{definition \textit{foldl}}\ \}{}\<[E]%
\\
\>[B]{}\hspace*{1cm}{}\<[BE]%
\>[6]{}\Varid{foldl}\;(\mathbin{+})\;(((\mathrm{0}\mathbin{+}\mathrm{3})\mathbin{+}\mathrm{7})\mathbin{+}\mathrm{5})\;[\mskip1.5mu \mskip1.5mu]{}\<[E]%
\\
\>[B]{}\mathrel{=}{}\<[BE]%
\>[8]{}\hspace*{5mm}\{\ \textrm{definition \textit{foldl}}\ \}{}\<[E]%
\\
\>[B]{}\hspace*{1cm}{}\<[BE]%
\>[6]{}((\mathrm{0}\mathbin{+}\mathrm{3})\mathbin{+}\mathrm{7})\mathbin{+}\mathrm{5}{}\<[E]%
\\
\>[B]{}\mathrel{=}{}\<[BE]%
\>[8]{}\hspace*{5mm}\{\ \textrm{applying $+$}\ \}{}\<[E]%
\\
\>[B]{}\hspace*{1cm}{}\<[BE]%
\>[6]{}(\mathrm{3}\mathbin{+}\mathrm{7})\mathbin{+}\mathrm{5}{}\<[E]%
\\
\>[B]{}\mathrel{=}{}\<[BE]%
\>[8]{}\hspace*{5mm}\{\ \textrm{applying $+$}\ \}{}\<[E]%
\\
\>[B]{}\hspace*{1cm}{}\<[BE]%
\>[6]{}\mathrm{10}\mathbin{+}\mathrm{5}{}\<[E]%
\\
\>[B]{}\mathrel{=}{}\<[BE]%
\>[8]{}\hspace*{5mm}\{\ \textrm{applying $+$}\ \}{}\<[E]%
\\
\>[B]{}\hspace*{1cm}{}\<[BE]%
\>[6]{}\mathrm{15}{}\<[E]%
\ColumnHook
\end{pboxed}
\)\par\noindent\endgroup\resethooks
\end{minipage}\quad
\begin{minipage}{7.5cm}
\begingroup\par\noindent\advance\leftskip\mathindent\(
\begin{pboxed}\SaveRestoreHook
\column{B}{@{}>{\hspre}c<{\hspost}@{}}%
\column{BE}{@{}l@{}}%
\column{6}{@{}>{\hspre}l<{\hspost}@{}}%
\column{8}{@{}>{\hspre}l<{\hspost}@{}}%
\column{E}{@{}>{\hspre}l<{\hspost}@{}}%
\>[B]{}\hspace*{1cm}{}\<[BE]%
\>[6]{}\Varid{sum}\;([\mskip1.5mu \mathrm{3},\mathrm{7}\mskip1.5mu]\plus [\mskip1.5mu \mathrm{5}\mskip1.5mu]){}\<[E]%
\\
\>[B]{}\mathrel{=}{}\<[BE]%
\>[8]{}\hspace*{5mm}\{\ \textrm{definition }\textit{sum}\ \}{}\<[E]%
\\
\>[B]{}\hspace*{1cm}{}\<[BE]%
\>[6]{}\Varid{foldl}\;(\mathbin{+})\;\mathrm{0}\;([\mskip1.5mu \mathrm{3},\mathrm{7}\mskip1.5mu]\plus [\mskip1.5mu \mathrm{5}\mskip1.5mu]){}\<[E]%
\\
\>[B]{}\mathrel{=}{}\<[BE]%
\>[8]{}\hspace*{5mm}\{\ \textrm{definition $+\!+$}\ \}{}\<[E]%
\\
\>[B]{}\hspace*{1cm}{}\<[BE]%
\>[6]{}\Varid{foldl}\;(\mathbin{+})\;\mathrm{0}\;(\mathrm{3}\mathbin{:}([\mskip1.5mu \mathrm{7}\mskip1.5mu]\plus [\mskip1.5mu \mathrm{5}\mskip1.5mu])){}\<[E]%
\\
\>[B]{}\mathrel{=}{}\<[BE]%
\>[8]{}\hspace*{5mm}\{\ \textrm{definition $+\!+$}\ \}{}\<[E]%
\\
\>[B]{}\hspace*{1cm}{}\<[BE]%
\>[6]{}\Varid{foldl}\;(\mathbin{+})\;\mathrm{0}\;(\mathrm{3}\mathbin{:}(\mathrm{7}\mathbin{:}([\mskip1.5mu \mskip1.5mu]\plus [\mskip1.5mu \mathrm{5}\mskip1.5mu]))){}\<[E]%
\\
\>[B]{}\mathrel{=}{}\<[BE]%
\>[8]{}\hspace*{5mm}\{\ \textrm{definition $+\!+$}\ \}{}\<[E]%
\\
\>[B]{}\hspace*{1cm}{}\<[BE]%
\>[6]{}\Varid{foldl}\;(\mathbin{+})\;\mathrm{0}\;[\mskip1.5mu \mathrm{3},\mathrm{7},\mathrm{5}\mskip1.5mu]{}\<[E]%
\\
\>[B]{}\mathrel{=}{}\<[BE]%
\>[8]{}\hspace*{5mm}\{\ \textrm{definition \textit{foldl}}\ \}{}\<[E]%
\\
\>[B]{}\hspace*{1cm}{}\<[BE]%
\>[6]{}\Varid{foldl}\;(\mathbin{+})\;(\mathrm{0}\mathbin{+}\mathrm{3})\;[\mskip1.5mu \mathrm{7},\mathrm{5}\mskip1.5mu]{}\<[E]%
\\
\>[B]{}\mathrel{=}{}\<[BE]%
\>[8]{}\hspace*{5mm}\{\ \textrm{applying $+$}\ \}{}\<[E]%
\\
\>[B]{}\hspace*{1cm}{}\<[BE]%
\>[6]{}\Varid{foldl}\;(\mathbin{+})\;\mathrm{3}\;[\mskip1.5mu \mathrm{7},\mathrm{5}\mskip1.5mu]{}\<[E]%
\\
\>[B]{}\mathrel{=}{}\<[BE]%
\>[8]{}\hspace*{5mm}\{\ \textrm{definition \textit{foldl}}\ \}{}\<[E]%
\\
\>[B]{}\hspace*{1cm}{}\<[BE]%
\>[6]{}\Varid{foldl}\;(\mathbin{+})\;(\mathrm{3}\mathbin{+}\mathrm{7})\;[\mskip1.5mu \mathrm{5}\mskip1.5mu]{}\<[E]%
\\
\>[B]{}\mathrel{=}{}\<[BE]%
\>[8]{}\hspace*{5mm}\{\ \textrm{applying $+$}\ \}{}\<[E]%
\\
\>[B]{}\hspace*{1cm}{}\<[BE]%
\>[6]{}\Varid{foldl}\;(\mathbin{+})\;\mathrm{10}\;[\mskip1.5mu \mathrm{5}\mskip1.5mu]{}\<[E]%
\\
\>[B]{}\mathrel{=}{}\<[BE]%
\>[8]{}\hspace*{5mm}\{\ \textrm{definition \textit{foldl}}\ \}{}\<[E]%
\\
\>[B]{}\hspace*{1cm}{}\<[BE]%
\>[6]{}\Varid{foldl}\;(\mathbin{+})\;(\mathrm{10}\mathbin{+}\mathrm{5})\;[\mskip1.5mu \mskip1.5mu]{}\<[E]%
\\
\>[B]{}\mathrel{=}{}\<[BE]%
\>[8]{}\hspace*{5mm}\{\ \textrm{applying $+$}\ \}{}\<[E]%
\\
\>[B]{}\hspace*{1cm}{}\<[BE]%
\>[6]{}\Varid{foldl}\;(\mathbin{+})\;\mathrm{15}\;[\mskip1.5mu \mskip1.5mu]{}\<[E]%
\\
\>[B]{}\mathrel{=}{}\<[BE]%
\>[8]{}\hspace*{5mm}\{\ \textrm{definition \textit{foldl}}\ \}{}\<[E]%
\\
\>[B]{}\hspace*{1cm}{}\<[BE]%
\>[6]{}\mathrm{15}{}\<[E]%
\ColumnHook
\end{pboxed}
\)\par\noindent\endgroup\resethooks
\end{minipage}
\end{displaymath}
\caption{Evaluating \ensuremath{\Varid{sum}\;([\mskip1.5mu \mathrm{3},\mathrm{7}\mskip1.5mu]\plus [\mskip1.5mu \mathrm{5}\mskip1.5mu])} using the outermost (left-hand side) or
innermost (right-hand side) evaluation strategy}
\label{fig:example}
\end{figure}

Figure~\ref{fig:example} shows two different ways to evaluate the expression 
\ensuremath{\Varid{sum}\;([\mskip1.5mu \mathrm{3},\mathrm{7}\mskip1.5mu]\plus [\mskip1.5mu \mathrm{5}\mskip1.5mu])}.
The evaluation steps on 
the left-hand side of Figure~\ref{fig:example} correspond to an outermost
evaluation order (call-by-name). After rewriting \ensuremath{\Varid{sum}} into a \ensuremath{\Varid{foldl}}, it 
is the list pattern in \ensuremath{\Varid{foldl}}'s definition (its third argument) that drives
evaluation. 
The evaluation steps nicely show the accumulating parameter of \ensuremath{\Varid{foldl}} for 
building up the result, the 
interleaving of steps for \ensuremath{\plus } (which produces a list) and \ensuremath{\Varid{foldl}} 
(which consumes list), and the additions that are calculated at the very end.
The evaluation steps on the right-hand side of Figure~\ref{fig:example}
illustrate the left-most innermost evaluation order (call-by-value), which 
fully evaluates sub-expression \ensuremath{[\mskip1.5mu \mathrm{3},\mathrm{7}\mskip1.5mu]\plus [\mskip1.5mu \mathrm{5}\mskip1.5mu]} before using \ensuremath{\Varid{foldl}}'s definition.
Observe that in contrast to
call-by-name evaluation the additions are immediately computed. Also observe that \ensuremath{\Varid{sum}} is immediately rewritten into \ensuremath{\Varid{foldl}}. 
This might be surprising behavior of an innermost evaluation strategy where 
arguments are completely evaluated before the function is evaluated. 
The reason for this behavior lies in the definition of \ensuremath{\Varid{sum}}. The definition 
of \ensuremath{\Varid{sum}} does not have an explicitly specified parameter, but it 
applies \ensuremath{\Varid{foldl}} partially. 
Therefore, the evaluator does not handle the sub-expression \ensuremath{[\mskip1.5mu \mathrm{3},\mathrm{7}\mskip1.5mu]\plus [\mskip1.5mu \mathrm{5}\mskip1.5mu]} as a 
child of \ensuremath{\Varid{sum}} but as a neighbor of \ensuremath{\Varid{sum}}. 

From an educational perspective it is
interesting to allow for alternative definitions of prelude functions, e.g.\
\ensuremath{\Varid{sum}} defined with explicit recursion, or \ensuremath{\Varid{sum}} defined with the strict \ensuremath{\Varid{foldl'}} function. With
our tool it is possible to switch between these alternative definitions and to
observe the consequences for evaluation.

It is important to keep in mind that the tool is capable of doing more than
only showing evaluation steps.
The tool also lets students practice with evaluating expressions, and can 
diagnose intermediate steps, suggest reducible expressions, and provide progress 
information by showing the number of evaluation steps remaining. It is possible
to train one particular evaluation strategy, or to allow any possible reduction step.

\section{Rewrite rules and strategies}
\label{section:rewriting}

We use \ideas{}, a framework for
developing domain reasoners that give intelligent feedback~\cite{rewritestrategiesforexercises}, for rewriting expressions. 
Many exercises, such as solving a mathematical equation or a programming exercise,
can be solved by following some kind of procedure. 
A procedure or strategy describes how basic steps may be 
combined to solve a particular
problem. Such a strategy is expressed in an
embedded domain-specific language. \ideas{} interprets a
strategy as a context-free grammar. The sentences of this grammar are
sequences of rewrite steps that are used to check if a student follows the
strategy. The main advantage of using \ideas{} in our tool is that it is a generic framework
that makes it possible to define exercises that must be solved using some kind
of strategy, and that it provides feedback to a student who is solving an
exercise. Feedback is added by means of labels at particular
locations in the strategy.

To use the \ideas{} framework, we construct three components:
the domain of the exercise (an expression datatype), rules for 
rewriting terms in this domain (the evaluation steps), and a rewrite strategy 
that combines these rules. Other components, such as parsing, pretty-printing, and 
testing expressions for equality, are omitted in this paper.

\begin{figure}[t]
\begingroup\par\noindent\advance\leftskip\mathindent\(
\begin{pboxed}\SaveRestoreHook
\column{B}{@{}>{\hspre}l<{\hspost}@{}}%
\column{12}{@{}>{\hspre}c<{\hspost}@{}}%
\column{12E}{@{}l@{}}%
\column{13}{@{}>{\hspre}l<{\hspost}@{}}%
\column{15}{@{}>{\hspre}l<{\hspost}@{}}%
\column{33}{@{}>{\hspre}l<{\hspost}@{}}%
\column{E}{@{}>{\hspre}l<{\hspost}@{}}%
\>[B]{}\mathbf{data}\;\Conid{Expr}{}\<[12]%
\>[12]{}\mathrel{=}{}\<[12E]%
\>[15]{}\Conid{App}\;\Conid{Expr}\;\Conid{Expr}{}\<[33]%
\>[33]{}\mbox{\onelinecomment  application}{}\<[E]%
\\
\>[12]{}\mid {}\<[12E]%
\>[15]{}\Conid{Abs}\;\Conid{String}\;\Conid{Expr}{}\<[33]%
\>[33]{}\mbox{\onelinecomment  lambda abstraction}{}\<[E]%
\\
\>[12]{}\mid {}\<[12E]%
\>[15]{}\Conid{Var}\;\Conid{String}{}\<[33]%
\>[33]{}\mbox{\onelinecomment  variable   }{}\<[E]%
\\
\>[12]{}\mid {}\<[12E]%
\>[15]{}\Conid{Lit}\;\Conid{Int}{}\<[33]%
\>[33]{}\mbox{\onelinecomment  integer            }{}\<[E]%
\\[\blanklineskip]%
\>[B]{}\mbox{\onelinecomment  smart constructors}{}\<[E]%
\\
\>[B]{}\Varid{appN}{}\<[13]%
\>[13]{}\mathrel{=}\Varid{foldl}\;\Conid{App}{}\<[33]%
\>[33]{}\mbox{\onelinecomment  n-ary application}{}\<[E]%
\\
\>[B]{}\Varid{nil}{}\<[13]%
\>[13]{}\mathrel{=}\Conid{Var}\;\text{\tt \char34 []\char34}{}\<[E]%
\\
\>[B]{}\Varid{cons}\;\Varid{x}\;\Varid{xs}{}\<[13]%
\>[13]{}\mathrel{=}\Varid{appN}\;(\Conid{Var}\;\text{\tt \char34 :\char34})\;[\mskip1.5mu \Varid{x},\Varid{xs}\mskip1.5mu]{}\<[E]%
\ColumnHook
\end{pboxed}
\)\par\noindent\endgroup\resethooks
\caption{Datatype for expressions}
\label{fig:expr}
\end{figure}
Figure~\ref{fig:expr} defines an expression datatype with application, lambda
abstraction, variables, and integers, together with some helper functions for 
constructing expressions. The \ensuremath{\Conid{Var}} constructor is also used to represent
datatype constructors (e.g., constructor \ensuremath{\mathbin{:}} for building lists). 

\subsection{Rewrite rules}
\label{section:rewriterules}
We introduce a rewrite rule for each function (and operator) from the prelude.
The rewrite rules are based on datatype-generic rewriting technology~\cite{JFP:rewriting}, 
where rules are constructed using operator \ensuremath{\leadsto}. This operator takes expressions on 
the left-hand side and the right-hand side. Based on \ensuremath{\Varid{sum}}'s definition, we 
define the rewrite rule for \ensuremath{\Varid{sum}} as follows:
\begin{spaced}\begingroup\par\noindent\advance\leftskip\mathindent\(
\begin{pboxed}\SaveRestoreHook
\column{B}{@{}>{\hspre}l<{\hspost}@{}}%
\column{4}{@{}>{\hspre}l<{\hspost}@{}}%
\column{7}{@{}>{\hspre}l<{\hspost}@{}}%
\column{E}{@{}>{\hspre}l<{\hspost}@{}}%
\>[B]{}\Varid{sumRule}\mathbin{::}\Conid{Rule}\;\Conid{Expr}{}\<[E]%
\\
\>[B]{}\Varid{sumRule}\mathrel{=}\Varid{describe}\;\text{\tt \char34 Calculate~the~sum~of~a~list~of~numbers\char34}\mathrel{\;\$\;}{}\<[E]%
\\
\>[B]{}\hsindent{4}{}\<[4]%
\>[4]{}\Varid{rewriteRule}\;\text{\tt \char34 eval.sum.rule\char34}\mathrel{\;\$\;}{}\<[E]%
\\
\>[4]{}\hsindent{3}{}\<[7]%
\>[7]{}\Conid{Var}\;\text{\tt \char34 sum\char34}\leadsto\Varid{appN}\;(\Conid{Var}\;\text{\tt \char34 foldl\char34})\;[\mskip1.5mu \Conid{Var}\;\text{\tt \char34 +\char34},\Conid{Lit}\;\mathrm{0}\mskip1.5mu]{}\<[E]%
\ColumnHook
\end{pboxed}
\)\par\noindent\endgroup\resethooks
\end{spaced}%
Each rule has an identifier (here \ensuremath{\text{\tt \char34 eval.sum.rule\char34}}) that is used for identifying
the rewrite step, and optionally also a description for explaining the step. 
The descriptions of the 
prelude functions are taken from the appendix of Hutton's textbook~\cite{Hutton}.
Note that functions such as \ensuremath{\Varid{describe}}, \ensuremath{\Varid{rewriteRule}}, operator \ensuremath{\leadsto} and type constructor
\ensuremath{\Conid{Rule}} are provided by the \ideas{} framework.

The rewrite rule for \ensuremath{\Varid{foldl}}'s
definition is more involved since it uses pattern matching. The pattern variables
in \ensuremath{\Varid{foldl}}'s definition are turned into meta-variables of the rewrite rule by
introducing these variables in a lambda abstraction:
\begin{spaced}\begingroup\par\noindent\advance\leftskip\mathindent\(
\begin{pboxed}\SaveRestoreHook
\column{B}{@{}>{\hspre}l<{\hspost}@{}}%
\column{3}{@{}>{\hspre}l<{\hspost}@{}}%
\column{5}{@{}>{\hspre}l<{\hspost}@{}}%
\column{21}{@{}>{\hspre}l<{\hspost}@{}}%
\column{24}{@{}>{\hspre}l<{\hspost}@{}}%
\column{E}{@{}>{\hspre}l<{\hspost}@{}}%
\>[B]{}\Varid{foldlRule}\mathbin{::}\Conid{Rule}\;\Conid{Expr}{}\<[E]%
\\
\>[B]{}\Varid{foldlRule}\mathrel{=}{}\<[E]%
\\
\>[B]{}\hsindent{3}{}\<[3]%
\>[3]{}\Varid{describe}\;\text{\tt \char34 Process~a~list~using~an~operator~that~associates~to~the~left\char34}\mathrel{\;\$\;}{}\<[E]%
\\
\>[B]{}\hsindent{3}{}\<[3]%
\>[3]{}\Varid{rewriteRules}\;\text{\tt \char34 eval.foldl.rule\char34}\;{}\<[E]%
\\
\>[3]{}\hsindent{2}{}\<[5]%
\>[5]{}[\mskip1.5mu \lambda \Varid{f}\;\Varid{v}\;\Varid{x}\;\Varid{xs}\to {}\<[21]%
\>[21]{}\Varid{appN}\;(\Conid{Var}\;\text{\tt \char34 foldl\char34})\;[\mskip1.5mu \Varid{f},\Varid{v},\Varid{nil}\mskip1.5mu]\leadsto\Varid{v}{}\<[E]%
\\
\>[3]{}\hsindent{2}{}\<[5]%
\>[5]{},\lambda \Varid{f}\;\Varid{v}\;\Varid{x}\;\Varid{xs}\to {}\<[21]%
\>[21]{}\Varid{appN}\;(\Conid{Var}\;\text{\tt \char34 foldl\char34})\;[\mskip1.5mu \Varid{f},\Varid{v},\Varid{cons}\;\Varid{x}\;\Varid{xs}\mskip1.5mu]\leadsto{}\<[E]%
\\
\>[21]{}\hsindent{3}{}\<[24]%
\>[24]{}\Varid{appN}\;(\Conid{Var}\;\text{\tt \char34 foldl\char34})\;[\mskip1.5mu \Varid{f},\Varid{appN}\;\Varid{f}\;[\mskip1.5mu \Varid{v},\Varid{x}\mskip1.5mu],\Varid{xs}\mskip1.5mu]{}\<[E]%
\\
\>[3]{}\hsindent{2}{}\<[5]%
\>[5]{}\mskip1.5mu]{}\<[E]%
\ColumnHook
\end{pboxed}
\)\par\noindent\endgroup\resethooks
\end{spaced}%

The rewrite rules \ensuremath{\Varid{sumRule}}, \ensuremath{\Varid{foldlRule}}, and \ensuremath{\Varid{appendRule}} (for operator \ensuremath{\plus }) 
have an intensional 
representation with a left- and right-hand 
side, which does not only make the rules easier to define, but also lets
us generate documentation for the rule, take the inverse of the rule, or
alter the matching algorithm for the rule's left-hand side (e.g., to take 
associativity of an operator into account).

Besides the rewrite rules, we also introduce a rule for the primitive 
addition function (\ensuremath{\Varid{addRule}}), and a rule for beta-reduction. Adding two 
integers cannot be defined by a rewrite rule, and therefore we use the function
\ensuremath{\Varid{makeRule}} from the \ideas{} framework to turn a function of type 
\ensuremath{\Conid{Expr}\to \Conid{Maybe}\;\Conid{Expr}} into a value of type \ensuremath{\Conid{Rule}\;\Conid{Expr}}.
\begin{spaced}\begingroup\par\noindent\advance\leftskip\mathindent\(
\begin{pboxed}\SaveRestoreHook
\column{B}{@{}>{\hspre}l<{\hspost}@{}}%
\column{3}{@{}>{\hspre}l<{\hspost}@{}}%
\column{5}{@{}>{\hspre}l<{\hspost}@{}}%
\column{46}{@{}>{\hspre}l<{\hspost}@{}}%
\column{E}{@{}>{\hspre}l<{\hspost}@{}}%
\>[B]{}\Varid{addRule}\mathbin{::}\Conid{Rule}\;\Conid{Expr}{}\<[E]%
\\
\>[B]{}\Varid{addRule}\mathrel{=}\Varid{describe}\;\text{\tt \char34 Add~two~integers\char34}\mathrel{\;\$\;}\Varid{makeRule}\;\text{\tt \char34 eval.add.rule\char34}\;\Varid{f}{}\<[E]%
\\
\>[B]{}\hsindent{3}{}\<[3]%
\>[3]{}\mathbf{where}{}\<[E]%
\\
\>[3]{}\hsindent{2}{}\<[5]%
\>[5]{}\Varid{f}\mathbin{::}\Conid{Expr}\to \Conid{Maybe}\;\Conid{Expr}{}\<[E]%
\\
\>[3]{}\hsindent{2}{}\<[5]%
\>[5]{}\Varid{f}\;(\Conid{App}\;(\Conid{App}\;(\Conid{Var}\;\text{\tt \char34 +\char34})\;(\Conid{Lit}\;\Varid{x}))\;(\Conid{Lit}\;\Varid{y})){}\<[46]%
\>[46]{}\mathrel{=}\Conid{Just}\mathrel{\;\$\;}\Conid{Lit}\;(\Varid{x}\mathbin{+}\Varid{y}){}\<[E]%
\\
\>[3]{}\hsindent{2}{}\<[5]%
\>[5]{}\Varid{f}\;\anonymous {}\<[46]%
\>[46]{}\mathrel{=}\Conid{Nothing}{}\<[E]%
\ColumnHook
\end{pboxed}
\)\par\noindent\endgroup\resethooks
\end{spaced}%
In a similar way we define \ensuremath{\Varid{betaReduction}\mathbin{::}\Conid{Rule}\;\Conid{Expr}} that reduces
expressions of the form \ensuremath{(\lambda \Varid{x}\to \Varid{e}_{1})\;\Varid{e}_{2}} by using \ensuremath{\Varid{makeRule}} and implementing 
a capture-avoiding substitution.

\subsection{Rewrite strategies}

\begin{figure}
\begin{center}
\begin{tabular}[t]{ll}
\hline
{\it Combinator} & {\it Description} \\[1mm]
\ensuremath{\Varid{s}\mathrel{{<}\hspace{-0.4em}\star\hspace{-0.4em}{>}}\Varid{t}} & first \ensuremath{\Varid{s}}, then \ensuremath{\Varid{t}} \\
\ensuremath{\Varid{s}\mathrel{{<}\hspace{-0.4em}\mid\hspace{-0.4em}{>}}\Varid{t}} & either \ensuremath{\Varid{s}} or \ensuremath{\Varid{t}} \\
\ensuremath{\Varid{s}\mathrel{\triangleright}\Varid{t}} & apply \ensuremath{\Varid{s}}, or else \ensuremath{\Varid{t}} \\
\ensuremath{\Varid{fix}\;\Varid{f}} & fixed point combinator \\
\ensuremath{\Varid{label}\;\ell\;\Varid{s}} & attach label \ensuremath{\ell} to \ensuremath{\Varid{s}} \\
\ensuremath{\Varid{succeed}} & always succeeds\\
\\
\hline
\end{tabular}\quad
\begin{tabular}[t]{ll}
\hline
{\it Combinator} & {\it Description} \\[1mm]
\ensuremath{\Varid{sequence}\;\Varid{xs}} & generalizes sequence (\ensuremath{\mathrel{{<}\hspace{-0.4em}\star\hspace{-0.4em}{>}}}) to lists \\
\ensuremath{\Varid{alternatives}\;\Varid{xs}} & generalizes choice (\ensuremath{\mathrel{{<}\hspace{-0.4em}\mid\hspace{-0.4em}{>}}}) to lists \\
\ensuremath{\Varid{repeat}\;\Varid{s}} & apply \ensuremath{\Varid{s}} as long as possible \\
\ensuremath{\Varid{child}\;\Varid{n}\;\Varid{s}} & apply \ensuremath{\Varid{s}} to the \ensuremath{\Varid{n}}-th child \\
\ensuremath{\Varid{outermost}\;\Varid{s}} & apply \ensuremath{\Varid{s}} at left-most outermost position\\
\ensuremath{\Varid{spinebu}\;\Varid{s}} & apply \ensuremath{\Varid{s}} to the left-spine (bottom-up) \\
\ensuremath{\Varid{checkCurrent}\;\Varid{p}} & succeeds if predicate \ensuremath{\Varid{p}} holds \\
\hline
\end{tabular}
\end{center}
\caption{Strategy combinators}
\label{fig:combinators}
\end{figure}
The embedded domain-specific language for specifying rewrite strategies in \ideas{} defines several generic combinators
to combine rewrite rules into a strategy~\cite{rewritestrategiesforexercises}.
We briefly introduce the combinators that are used in this paper. The 
combinators are summarized in Figure~\ref{fig:combinators}.

Rewrite rules are the basic building block for composing rewrite strategies. 
All strategy combinators in the \ideas{} framework 
are overloaded and take rules or strategies as arguments. The sequence
combinator (\ensuremath{\mathrel{{<}\hspace{-0.4em}\star\hspace{-0.4em}{>}}}) specifies the sequential application of two strategies. The
choice combinator (\ensuremath{\mathrel{{<}\hspace{-0.4em}\mid\hspace{-0.4em}{>}}}) defines that either the first operand or the second
operand is applied: combinator \ensuremath{\Varid{alternatives}} generalizes the choice
combinator to lists. The left-biased choice combinator (\ensuremath{\mathrel{\triangleright}}) only tries
the second strategy if the first strategy fails.
Combinator \ensuremath{\Varid{repeat}} is used for repetition: this
combinator applies its argument strategy as often as possible. The fixed point
combinator \ensuremath{\Varid{fix}} is used to explicitly model recursion in the strategy. It takes
as argument a function that maps a strategy to a new strategy.
We can use labels at any position in the strategy to specialize the feedback
that is generated. 

The strategy language supports all the usual traversal
combinators such as \ensuremath{\Varid{innermost}} and \ensuremath{\Varid{oncebu}}~\cite{VBT98}.
The strategy \ensuremath{\Varid{child}\;\Varid{n}\;\Varid{s}} applies strategy \ensuremath{\Varid{s}} to the \ensuremath{\Varid{n}}-th child and can 
be used to define other generic traversal combinators.
Combinator \ensuremath{\Varid{checkCurrent}} takes a predicate and only succeeds if
the predicate holds for the current expression. 

An evaluation strategy defines in which order sub-expressions are reduced.
We can use the standard left-most outermost (or innermost)  rewrite 
strategy to turn the rewrite rules into an evaluation
strategy:
\begin{spaced}\begingroup\par\noindent\advance\leftskip\mathindent\(
\begin{pboxed}\SaveRestoreHook
\column{B}{@{}>{\hspre}l<{\hspost}@{}}%
\column{3}{@{}>{\hspre}l<{\hspost}@{}}%
\column{E}{@{}>{\hspre}l<{\hspost}@{}}%
\>[B]{}\Varid{rules}\mathbin{::}[\mskip1.5mu \Conid{Rule}\;\Conid{Expr}\mskip1.5mu]{}\<[E]%
\\
\>[B]{}\Varid{rules}\mathrel{=}[\mskip1.5mu \Varid{sumRule},\Varid{foldlRule},\Varid{appendRule},\Varid{addRule},\Varid{betaReduction}\mskip1.5mu]{}\<[E]%
\\[\blanklineskip]%
\>[B]{}\Varid{evalOutermost}\mathbin{::}\Conid{LabeledStrategy}\;(\Conid{Context}\;\Conid{Expr}){}\<[E]%
\\
\>[B]{}\Varid{evalOutermost}\mathrel{=}\Varid{label}\;\text{\tt \char34 eval.outer\char34}\mathrel{\;\$\;}{}\<[E]%
\\
\>[B]{}\hsindent{3}{}\<[3]%
\>[3]{}\Varid{outermost}\;(\Varid{alternatives}\;(\Varid{map}\;\Varid{liftToContext}\;\Varid{rules})){}\<[E]%
\ColumnHook
\end{pboxed}
\)\par\noindent\endgroup\resethooks
\end{spaced}%
Note that we use the \ensuremath{\Conid{Context}} type from the \ideas{} framework as a 
zipper~\cite{HuetZipper} datatype for traversing expressions. A zipper maintains
a sub-expression that has the focus, and is used by traversal combinators such as 
\ensuremath{\Varid{outermost}}. We lift the rules to the \ensuremath{\Conid{Context}} type with the function
\ensuremath{\Varid{liftToContext}\mathbin{::}\Conid{Rule}\;\Varid{a}\to \Conid{Rule}\;(\Conid{Context}\;\Varid{a})} that applies the rule
to the sub-expression that currently has the focus.

The attentive reader will have noticed that the \ensuremath{\Varid{evalOutermost}} strategy
does not result in the evaluation that is shown on the left-hand side of 
Figure~\ref{fig:example}.
Evaluation of a \ensuremath{\Varid{foldl}} application is driven by pattern matching on the 
function's third argument (the list). The expression at this position should 
first be evaluated to weak-head normal form (whnf), after which we can decide which
case to take. We define a rewrite strategy that first checks that \ensuremath{\Varid{foldl}} is
applied to exactly three arguments, then brings the third argument to 
weak-head normal form, and finally applies the rewrite rule for \ensuremath{\Varid{foldl}}. 
We need a strategy that can evaluate an expression to weak-head normal form, 
and pass this as an argument to the strategy definition.
\begin{spaced}\begingroup\par\noindent\advance\leftskip\mathindent\(
\begin{pboxed}\SaveRestoreHook
\column{B}{@{}>{\hspre}l<{\hspost}@{}}%
\column{4}{@{}>{\hspre}c<{\hspost}@{}}%
\column{4E}{@{}l@{}}%
\column{9}{@{}>{\hspre}l<{\hspost}@{}}%
\column{42}{@{}>{\hspre}l<{\hspost}@{}}%
\column{E}{@{}>{\hspre}l<{\hspost}@{}}%
\>[B]{}\Varid{foldlS}\mathbin{::}\Conid{Strategy}\;(\Conid{Context}\;\Conid{Expr})\to \Conid{LabeledStrategy}\;(\Conid{Context}\;\Conid{Expr}){}\<[E]%
\\
\>[B]{}\Varid{foldlS}\;\Varid{whnf}\mathrel{=}\Varid{label}\;\text{\tt \char34 eval.foldl\char34}\mathrel{\;\$\;}{}\<[E]%
\\
\>[B]{}\hsindent{9}{}\<[9]%
\>[9]{}\Varid{checkCurrent}\;(\Varid{isFun}\;\text{\tt \char34 foldl\char34}\;\mathrm{3}){}\<[42]%
\>[42]{}\mbox{\onelinecomment  check that \ensuremath{\Varid{foldl}} has exactly 3 arguments}{}\<[E]%
\\
\>[B]{}\hsindent{4}{}\<[4]%
\>[4]{}\mathrel{{<}\hspace{-0.4em}\star\hspace{-0.4em}{>}}{}\<[4E]%
\>[9]{}\Varid{arg}\;\mathrm{3}\;\mathrm{3}\;\Varid{whnf}{}\<[42]%
\>[42]{}\mbox{\onelinecomment  bring the third argument (out of 3) to whnf}{}\<[E]%
\\
\>[B]{}\hsindent{4}{}\<[4]%
\>[4]{}\mathrel{{<}\hspace{-0.4em}\star\hspace{-0.4em}{>}}{}\<[4E]%
\>[9]{}\Varid{liftToContext}\;\Varid{foldlRule}{}\<[42]%
\>[42]{}\mbox{\onelinecomment  apply the rewrite rule for \ensuremath{\Varid{foldl}}'s definition}{}\<[E]%
\ColumnHook
\end{pboxed}
\)\par\noindent\endgroup\resethooks
\end{spaced}%
\begin{figure}
\begingroup\par\noindent\advance\leftskip\mathindent\(
\begin{pboxed}\SaveRestoreHook
\column{B}{@{}>{\hspre}l<{\hspost}@{}}%
\column{11}{@{}>{\hspre}l<{\hspost}@{}}%
\column{14}{@{}>{\hspre}l<{\hspost}@{}}%
\column{18}{@{}>{\hspre}l<{\hspost}@{}}%
\column{25}{@{}>{\hspre}l<{\hspost}@{}}%
\column{E}{@{}>{\hspre}l<{\hspost}@{}}%
\>[B]{}\Varid{isApp}\mathbin{::}\Conid{Expr}\to \Conid{Bool}{}\<[E]%
\\
\>[B]{}\Varid{isApp}\;(\Conid{App}\;\anonymous \;\anonymous ){}\<[18]%
\>[18]{}\mathrel{=}\Conid{True}{}\<[E]%
\\
\>[B]{}\Varid{isApp}\;\anonymous {}\<[18]%
\>[18]{}\mathrel{=}\Conid{False}{}\<[E]%
\\[\blanklineskip]%
\>[B]{}\Varid{isFun}\mathbin{::}\Conid{String}\to \Conid{Int}\to \Conid{Expr}\to \Conid{Bool}{}\<[E]%
\\
\>[B]{}\Varid{isFun}\;\Varid{fn}\;{}\<[11]%
\>[11]{}\mathrm{0}\;{}\<[14]%
\>[14]{}(\Conid{Var}\;\Varid{s}){}\<[25]%
\>[25]{}\mathrel{=}\Varid{fn}\mathbin{==}\Varid{s}{}\<[E]%
\\
\>[B]{}\Varid{isFun}\;\Varid{fn}\;{}\<[11]%
\>[11]{}\Varid{n}\;{}\<[14]%
\>[14]{}(\Conid{App}\;\Varid{f}\;\anonymous ){}\<[25]%
\>[25]{}\mathrel{=}\Varid{isFun}\;\Varid{fn}\;(\Varid{n}\mathbin{-}\mathrm{1})\;\Varid{f}{}\<[E]%
\\
\>[B]{}\Varid{isFun}\;\anonymous \;{}\<[11]%
\>[11]{}\anonymous \;{}\<[14]%
\>[14]{}\anonymous {}\<[25]%
\>[25]{}\mathrel{=}\Conid{False}{}\<[E]%
\ColumnHook
\end{pboxed}
\)\par\noindent\endgroup\resethooks
\caption{Predicates on expressions (helper functions)}
\label{fig:expr-predicates}
\end{figure}%
The predicate \ensuremath{\Varid{isFun}} (defined in Figure~\ref{fig:expr-predicates}) tests
whether an expression is a function application of a specific function with an 
exact number of arguments. We define the strategy combinator \ensuremath{\Varid{arg}\;\Varid{i}\;\Varid{n}\;\Varid{s}}, used 
in the definition of \ensuremath{\Varid{foldlS}}, to apply strategy \ensuremath{\Varid{s}} to the \ensuremath{\Varid{i}}-th argument of
a function application with \ensuremath{\Varid{n}} arguments. For the last argument (\ensuremath{\Varid{i}\mathbin{==}\Varid{n}}), 
we apply \ensuremath{\Varid{s}} to the second sub-expression of an application. Otherwise, we 
visit the first sub-expression and call \ensuremath{\Varid{arg}} recursively. The combinator is
not defined for \ensuremath{\Varid{i}\mathbin{>}\Varid{n}}.
\begin{spaced}\begingroup\par\noindent\advance\leftskip\mathindent\(
\begin{pboxed}\SaveRestoreHook
\column{B}{@{}>{\hspre}l<{\hspost}@{}}%
\column{12}{@{}>{\hspre}l<{\hspost}@{}}%
\column{22}{@{}>{\hspre}l<{\hspost}@{}}%
\column{E}{@{}>{\hspre}l<{\hspost}@{}}%
\>[B]{}\Varid{arg}\mathbin{::}\Conid{Int}\to \Conid{Int}\to \Conid{Strategy}\;(\Conid{Context}\;\Varid{a})\to \Conid{Strategy}\;(\Conid{Context}\;\Varid{a}){}\<[E]%
\\
\>[B]{}\Varid{arg}\;\Varid{i}\;\Varid{n}\;\Varid{s}{}\<[12]%
\>[12]{}\mid \Varid{i}\mathbin{==}\Varid{n}{}\<[22]%
\>[22]{}\mathrel{=}\Varid{child}\;\mathrm{1}\;\Varid{s}{}\<[E]%
\\
\>[12]{}\mid \Varid{i}\mathbin{<}\Varid{n}{}\<[22]%
\>[22]{}\mathrel{=}\Varid{child}\;\mathrm{0}\;(\Varid{arg}\;\Varid{i}\;(\Varid{n}\mathbin{-}\mathrm{1})\;\Varid{s}){}\<[E]%
\ColumnHook
\end{pboxed}
\)\par\noindent\endgroup\resethooks
\end{spaced}%

We also give the evaluation strategy for the definition of the append 
function, to emphasize its similarity with the definition of \ensuremath{\Varid{foldlS}}.
Other function definitions have a similar evaluation strategy.

\begin{spaced}\begingroup\par\noindent\advance\leftskip\mathindent\(
\begin{pboxed}\SaveRestoreHook
\column{B}{@{}>{\hspre}l<{\hspost}@{}}%
\column{4}{@{}>{\hspre}c<{\hspost}@{}}%
\column{4E}{@{}l@{}}%
\column{9}{@{}>{\hspre}l<{\hspost}@{}}%
\column{38}{@{}>{\hspre}l<{\hspost}@{}}%
\column{E}{@{}>{\hspre}l<{\hspost}@{}}%
\>[B]{}\Varid{appendS}\mathbin{::}\Conid{Strategy}\;(\Conid{Context}\;\Conid{Expr})\to \Conid{LabeledStrategy}\;(\Conid{Context}\;\Conid{Expr}){}\<[E]%
\\
\>[B]{}\Varid{appendS}\;\Varid{whnf}\mathrel{=}\Varid{label}\;\text{\tt \char34 eval.append\char34}\mathrel{\;\$\;}{}\<[E]%
\\
\>[B]{}\hsindent{9}{}\<[9]%
\>[9]{}\Varid{checkCurrent}\;(\Varid{isFun}\;\text{\tt \char34 ++\char34}\;\mathrm{2}){}\<[38]%
\>[38]{}\mbox{\onelinecomment  check that append (\ensuremath{\plus }) has exactly 2 arguments}{}\<[E]%
\\
\>[B]{}\hsindent{4}{}\<[4]%
\>[4]{}\mathrel{{<}\hspace{-0.4em}\star\hspace{-0.4em}{>}}{}\<[4E]%
\>[9]{}\Varid{arg}\;\mathrm{1}\;\mathrm{2}\;\Varid{whnf}{}\<[38]%
\>[38]{}\mbox{\onelinecomment  bring the first argument (out of 2) to whnf}{}\<[E]%
\\
\>[B]{}\hsindent{4}{}\<[4]%
\>[4]{}\mathrel{{<}\hspace{-0.4em}\star\hspace{-0.4em}{>}}{}\<[4E]%
\>[9]{}\Varid{liftToContext}\;\Varid{appendRule}{}\<[38]%
\>[38]{}\mbox{\onelinecomment  apply the rewrite rule for append's definition}{}\<[E]%
\ColumnHook
\end{pboxed}
\)\par\noindent\endgroup\resethooks
\end{spaced}%

Evaluating an expression to weak-head normal form is a fixed-point computation
over the evaluation strategies for the definitions, since each definition takes
the \ensuremath{\Varid{whnf}} strategy as an argument. We combine the evaluation strategies with the 
rewrite rule for beta-reduction, apply it to the left-spine of an application
(in a bottom-up way), and repeat this until the strategy can no longer be 
applied. This brings the expression in weak-head normal form.
\begin{spaced}\begingroup\par\noindent\advance\leftskip\mathindent\(
\begin{pboxed}\SaveRestoreHook
\column{B}{@{}>{\hspre}l<{\hspost}@{}}%
\column{4}{@{}>{\hspre}l<{\hspost}@{}}%
\column{E}{@{}>{\hspre}l<{\hspost}@{}}%
\>[B]{}\Varid{prelude}\mathbin{::}[\mskip1.5mu \Conid{Strategy}\;(\Conid{Context}\;\Conid{Expr})\to \Conid{LabeledStrategy}\;(\Conid{Context}\;\Conid{Expr})\mskip1.5mu]{}\<[E]%
\\
\>[B]{}\Varid{prelude}\mathrel{=}[\mskip1.5mu \Varid{sumS},\Varid{foldlS},\Varid{appendS},\Varid{addS}\mskip1.5mu]{}\<[E]%
\\[\blanklineskip]%
\>[B]{}\Varid{spinebu}\mathbin{::}\Conid{Strategy}\;(\Conid{Context}\;\Conid{Expr})\to \Conid{Strategy}\;(\Conid{Context}\;\Conid{Expr}){}\<[E]%
\\
\>[B]{}\Varid{spinebu}\;\Varid{s}\mathrel{=}\Varid{fix}\mathrel{\;\$\;}\lambda \Varid{x}\to (\Varid{checkCurrent}\;\Varid{isApp}\mathrel{{<}\hspace{-0.4em}\star\hspace{-0.4em}{>}}\Varid{child}\;\mathrm{0}\;\Varid{x})\mathrel{\triangleright}\Varid{s}{}\<[E]%
\\[\blanklineskip]%
\>[B]{}\Varid{evalWhnf}\mathbin{::}\Conid{LabeledStrategy}\;(\Conid{Context}\;\Conid{Expr}){}\<[E]%
\\
\>[B]{}\Varid{evalWhnf}\mathrel{=}\Varid{label}\;\text{\tt \char34 eval.whnf\char34}\mathrel{\;\$\;}\Varid{fix}\mathrel{\;\$\;}\lambda \Varid{whnf}\to {}\<[E]%
\\
\>[B]{}\hsindent{4}{}\<[4]%
\>[4]{}\Varid{repeat}\;(\Varid{spinebu}\;(\Varid{liftToContext}\;\Varid{betaReduction}\mathrel{{<}\hspace{-0.4em}\mid\hspace{-0.4em}{>}}\Varid{alternatives}\;[\mskip1.5mu \Varid{f}\;\Varid{whnf}\mid \Varid{f}\leftarrow \Varid{prelude}\mskip1.5mu])){}\<[E]%
\ColumnHook
\end{pboxed}
\)\par\noindent\endgroup\resethooks
\end{spaced}%
The \ensuremath{\Varid{repeat}} combinator in the definition of \ensuremath{\Varid{evalWhnf}} applies its argument
strategy zero or more times. For example, the strategy does not have to be 
applied for an expression that already is in weak-head normal form, and is 
applied twice to rewrite the expression \ensuremath{(\Varid{id}\;\Varid{id})\;\mathrm{3}} into \ensuremath{\Varid{id}\;\mathrm{3}} 
and then \ensuremath{\mathrm{3}} (where \ensuremath{\Varid{id}} is Haskell's identity function).

The result of applying \ensuremath{\Varid{evalWhnf}} to the expression \ensuremath{\Varid{sum}\;([\mskip1.5mu \mathrm{3},\mathrm{7}\mskip1.5mu]\plus [\mskip1.5mu \mathrm{5}\mskip1.5mu])} gives
the calculation shown in Figure~\ref{fig:example} on the left-hand side. To fully evaluate 
a value (such as a list), we have to repeat the \ensuremath{\Varid{evalWhnf}} strategy for the sub-parts
of a constructor.

\subsection{User-defined function definitions}
From an educational viewpoint it is interesting to experiment with different function definitions 
for the same function and to observe the implications. For example, the function \ensuremath{\Varid{sum}} can be defined 
using \ensuremath{\Varid{foldl}}, \ensuremath{\Varid{foldr}}, the strict \ensuremath{\Varid{foldl'}}, or as an explicit recursive definition. 
This feature is also interesting for instructors 
who would like to incorporate the tool into their course on functional programming because they 
can then use their own examples to explain programming concepts to their students.  

Multiple functions definitions can be added by giving each definition a different name 
(for instance, \ensuremath{\Varid{sum}}, \ensuremath{\Varid{sum'}}, and \ensuremath{\Varid{sum''}}) and by manually adding rewrite rules and evaluation strategies 
for these definitions. However, this approach has some drawbacks. To define rewrite rules and evaluation 
strategies, the maintainer of the prototype (probably the instructor) needs to have knowledge of the datatype for Haskell expressions and of the internal implementation to construct rules and 
evaluation strategies. After adding definitions, the prototype needs to be recompiled, and hence a maintainer 
also needs a complete build environment (Haskell compiler and libraries used). Another 
disadvantage is that the translation of function definitions to rewrite rules is a manual, error-prone process. 

Function definitions have rewrite rules and evaluation strategies of a similar structure. 
An alternative approach to manually adding rewriting strategies is therefore to generate rewrite rules and evaluation strategies from function definitions. These function definitions can be defined in a Haskell source file, and using
annotations~\cite{gerdes2012ask} we can add a description to every function definition. 
Annotations are multi-line comments, so the file remains a valid Haskell 
source file. Annotations have an identifier, for instance DESC for description, which is used by the prototype to interpret 
the string after DESC as the rule description. With this approach there is 
a single file, located on the web server, that contains the function definitions, and their corresponding 
descriptions. No knowledge of the tool's internal implementation is required to
add or change function definitions. 
An example of such a file is given below:

\begin{spaced}\begingroup\par\noindent\advance\leftskip\mathindent\(
\begin{pboxed}\SaveRestoreHook
\column{B}{@{}>{\hspre}l<{\hspost}@{}}%
\column{17}{@{}>{\hspre}l<{\hspost}@{}}%
\column{E}{@{}>{\hspre}l<{\hspost}@{}}%
\>[B]{}\mbox{\enskip\{-\# DESC \ensuremath{\Varid{sum}} defined with a \ensuremath{\Varid{foldr}} to sum up all elements of a list.  \#-\}\enskip}{}\<[E]%
\\
\>[B]{}\Varid{sum'}\mathrel{=}\Varid{foldr}\;(\mathbin{+})\;\mathrm{0}{}\<[E]%
\\[\blanklineskip]%
\>[B]{}\mbox{\enskip\{-\# DESC \ensuremath{\Varid{sum}} defined recursively to sum up all elements of a list.  \#-\}\enskip}{}\<[E]%
\\
\>[B]{}\Varid{sum''}\;[\mskip1.5mu \mskip1.5mu]{}\<[17]%
\>[17]{}\mathrel{=}\mathrm{0}{}\<[E]%
\\
\>[B]{}\Varid{sum''}\;(\Varid{x}\mathbin{:}\Varid{xs}){}\<[17]%
\>[17]{}\mathrel{=}\Varid{x}\mathbin{+}\Varid{sum''}\;\Varid{xs}{}\<[E]%
\\[\blanklineskip]%
\>[B]{}\mbox{\enskip\{-\# DESC \ensuremath{\Varid{double}} function to double a number.  \#-\}\enskip}{}\<[E]%
\\
\>[B]{}\Varid{double}\;\Varid{x}\mathrel{=}\Varid{x}\mathbin{+}\Varid{x}{}\<[E]%
\ColumnHook
\end{pboxed}
\)\par\noindent\endgroup\resethooks
\end{spaced}%

We name this file the evaluator's \emph{prelude} because it defines the 
functions that can be used in the evaluator.
Notice that primitive functions (such as the operator \ensuremath{\mathbin{+}}) cannot be defined in 
the prelude file. Primitive functions are functions that cannot be 
implemented directly in Haskell and are provided natively by the compiler. 

The function definitions given in the example prelude file use
pattern matching. The left-hand side of the equal sign consists of the 
function name and zero or more patterns. If the function name and the 
patterns match with a particular expression the expression is rewritten to the
expression at the right-hand side of the equal sign (after substituting the
pattern variables). Figure~\ref{fig:patterns} defines
a datatype for patterns and function definitions.

\begin{figure}
\begingroup\par\noindent\advance\leftskip\mathindent\(
\begin{pboxed}\SaveRestoreHook
\column{B}{@{}>{\hspre}l<{\hspost}@{}}%
\column{11}{@{}>{\hspre}c<{\hspost}@{}}%
\column{11E}{@{}l@{}}%
\column{14}{@{}>{\hspre}l<{\hspost}@{}}%
\column{50}{@{}>{\hspre}l<{\hspost}@{}}%
\column{E}{@{}>{\hspre}l<{\hspost}@{}}%
\>[B]{}\mathbf{data}\;\Conid{Pat}{}\<[11]%
\>[11]{}\mathrel{=}{}\<[11E]%
\>[14]{}\Conid{PCon}\;\Conid{String}\;[\mskip1.5mu \Conid{Pat}\mskip1.5mu]{}\<[50]%
\>[50]{}\mbox{\onelinecomment  pattern constructor}{}\<[E]%
\\
\>[11]{}\mid {}\<[11E]%
\>[14]{}\Conid{PVar}\;\Conid{String}{}\<[50]%
\>[50]{}\mbox{\onelinecomment  pattern variable}{}\<[E]%
\\
\>[11]{}\mid {}\<[11E]%
\>[14]{}\Conid{PLit}\;\Conid{Int}{}\<[50]%
\>[50]{}\mbox{\onelinecomment  integer pattern  }{}\<[E]%
\\[\blanklineskip]%
\>[B]{}\mathbf{data}\;\Conid{Def}{}\<[11]%
\>[11]{}\mathrel{=}{}\<[11E]%
\>[14]{}\Conid{Def}\;\Conid{String}\;[\mskip1.5mu \Conid{Pat}\mskip1.5mu]\;\Conid{Expr}{}\<[50]%
\>[50]{}\mbox{\onelinecomment  function definition}{}\<[E]%
\ColumnHook
\end{pboxed}
\)\par\noindent\endgroup\resethooks
\caption{Datatype for patterns and function definitions}
\label{fig:patterns}
\end{figure}

A function definition can have multiple function bindings. For example,
the function \ensuremath{\Varid{foldl}} has a binding for the empty list and the non-empty list.
In the remainder of the paper we will only consider function definitions with
one binding. Definitions with more bindings can be supported by combining 
generated rules and strategies. An evaluation strategy is generated for 
each function binding, and these strategies are tried in order of appearance in
the corresponding definition. For this, we use the strategy combinator 
\ensuremath{\mathrel{\triangleright}} from the \ideas{} framework.

\paragraph{Generation of rewrite rules.} 

We need a function that takes a rule identifier (that may contain a description) 
and a function definition and turns these into a rewrite rule for expressions.
\begin{spaced}\begingroup\par\noindent\advance\leftskip\mathindent\(
\begin{pboxed}\SaveRestoreHook
\column{B}{@{}>{\hspre}l<{\hspost}@{}}%
\column{E}{@{}>{\hspre}l<{\hspost}@{}}%
\>[B]{}\Varid{genRule}\mathbin{::}\Conid{Id}\to \Conid{Def}\to \Conid{Rule}\;\Conid{Expr}{}\<[E]%
\ColumnHook
\end{pboxed}
\)\par\noindent\endgroup\resethooks
\end{spaced}%
Rewrite rules are constructed with operator \ensuremath{\leadsto} that expects two expressions
as its operands. Hence, we need a function to convert the patterns on the 
left-hand side of a definition into an expression.
\begin{spaced}\begingroup\par\noindent\advance\leftskip\mathindent\(
\begin{pboxed}\SaveRestoreHook
\column{B}{@{}>{\hspre}l<{\hspost}@{}}%
\column{24}{@{}>{\hspre}l<{\hspost}@{}}%
\column{E}{@{}>{\hspre}l<{\hspost}@{}}%
\>[B]{}\Varid{patToExpr}\mathbin{::}\Conid{Pat}\to \Conid{Expr}{}\<[E]%
\\
\>[B]{}\Varid{patToExpr}\;(\Conid{PCon}\;\Varid{s}\;\Varid{ps}){}\<[24]%
\>[24]{}\mathrel{=}\Varid{appN}\;(\Conid{Var}\;\Varid{s})\;(\Varid{map}\;\Varid{patToExpr}\;\Varid{ps}){}\<[E]%
\\
\>[B]{}\Varid{patToExpr}\;(\Conid{PVar}\;\Varid{s}){}\<[24]%
\>[24]{}\mathrel{=}\Conid{Var}\;\Varid{s}{}\<[E]%
\\
\>[B]{}\Varid{patToExpr}\;(\Conid{PLit}\;\Varid{n}){}\<[24]%
\>[24]{}\mathrel{=}\Conid{Lit}\;\Varid{n}{}\<[E]%
\ColumnHook
\end{pboxed}
\)\par\noindent\endgroup\resethooks
\end{spaced}%
Note that the pattern variables of a function definition act as meta-variables 
in the rewrite rule, and the corresponding variables on the right-hand side 
have to be substituted when the rewrite rule is applied. We omit the 
definition of \ensuremath{\Varid{genRule}} that uses \ensuremath{\Varid{patToExpr}}.

\paragraph{Generation of evaluation strategies.} 

We can generate an evaluation strategy for a function definition used in the
outermost evaluation strategy based on pattern matching. Such an evaluation 
strategy consists of three steps: check the name of the function, bring some
arguments to weak-head normal form, and apply the definition's rewrite rule.
We first focus on the second step of bringing arguments to weak-head normal form
and define strategy combinator \ensuremath{\Varid{args}} for applying a list of strategies
to the arguments of a function application. We reuse combinator \ensuremath{\Varid{arg}}:
\begin{spaced}\begingroup\par\noindent\advance\leftskip\mathindent\(
\begin{pboxed}\SaveRestoreHook
\column{B}{@{}>{\hspre}l<{\hspost}@{}}%
\column{E}{@{}>{\hspre}l<{\hspost}@{}}%
\>[B]{}\Varid{args}\mathbin{::}[\mskip1.5mu \Conid{Strategy}\;(\Conid{Context}\;\Varid{a})\mskip1.5mu]\to \Conid{Strategy}\;(\Conid{Context}\;\Varid{a}){}\<[E]%
\\
\>[B]{}\Varid{args}\;\Varid{xs}\mathrel{=}\Varid{sequence}\;[\mskip1.5mu \Varid{arg}\;\Varid{i}\;(\Varid{length}\;\Varid{xs})\;\Varid{s}\mid (\Varid{i},\Varid{s})\leftarrow \Varid{zip}\;[\mskip1.5mu \mathrm{1}\mathinner{\ldotp\ldotp}\mskip1.5mu]\;\Varid{xs}\mskip1.5mu]{}\<[E]%
\ColumnHook
\end{pboxed}
\)\par\noindent\endgroup\resethooks
\end{spaced}%
The function \ensuremath{\Varid{patS}} constructs an evaluation strategy for a given pattern. For
pattern variables, nothing has to be evaluated. For a pattern constructor, we
bring the expression to weak-head normal form, check the name of the 
constructor, and then recursively deal with the constructor's sub-patterns.
For a literal pattern, we bring the expression to weak-head normal form
and then check whether it is the same number.
\begin{spaced}\begingroup\par\noindent\advance\leftskip\mathindent\(
\begin{pboxed}\SaveRestoreHook
\column{B}{@{}>{\hspre}l<{\hspost}@{}}%
\column{21}{@{}>{\hspre}c<{\hspost}@{}}%
\column{21E}{@{}l@{}}%
\column{24}{@{}>{\hspre}l<{\hspost}@{}}%
\column{26}{@{}>{\hspre}l<{\hspost}@{}}%
\column{E}{@{}>{\hspre}l<{\hspost}@{}}%
\>[B]{}\Varid{patS}\mathbin{::}\Conid{Strategy}\;(\Conid{Context}\;\Conid{Expr})\to \Conid{Pat}\to \Conid{Strategy}\;(\Conid{Context}\;\Conid{Expr}){}\<[E]%
\\
\>[B]{}\Varid{patS}\;\Varid{whnf}\;(\Conid{PCon}\;\Varid{s}\;\Varid{ps}){}\<[24]%
\>[24]{}\mathrel{=}\Varid{whnf}\mathrel{{<}\hspace{-0.4em}\star\hspace{-0.4em}{>}}\Varid{patListS}\;\Varid{whnf}\;\Varid{s}\;\Varid{ps}{}\<[E]%
\\
\>[B]{}\Varid{patS}\;\Varid{whnf}\;(\Conid{PVar}\;\Varid{s}){}\<[24]%
\>[24]{}\mathrel{=}\Varid{succeed}{}\<[E]%
\\
\>[B]{}\Varid{patS}\;\Varid{whnf}\;(\Conid{PLit}\;\Varid{n}){}\<[24]%
\>[24]{}\mathrel{=}\Varid{whnf}\mathrel{{<}\hspace{-0.4em}\star\hspace{-0.4em}{>}}\Varid{checkCurrent}\;(\mathbin{==}\Conid{Lit}\;\Varid{n}){}\<[E]%
\\[\blanklineskip]%
\>[B]{}\Varid{patListS}\mathbin{::}\Conid{Strategy}\;(\Conid{Context}\;\Conid{Expr})\to \Conid{String}\to [\mskip1.5mu \Conid{Pat}\mskip1.5mu]\to \Conid{Strategy}\;(\Conid{Context}\;\Conid{Expr}){}\<[E]%
\\
\>[B]{}\Varid{patListS}\;\Varid{whnf}\;\Varid{s}\;\Varid{ps}{}\<[21]%
\>[21]{}\mathrel{=}{}\<[21E]%
\>[26]{}\Varid{checkCurrent}\;(\Varid{isFun}\;\Varid{s}\;(\Varid{length}\;\Varid{ps})){}\<[E]%
\\
\>[21]{}\mathrel{{<}\hspace{-0.4em}\star\hspace{-0.4em}{>}}{}\<[21E]%
\>[26]{}\Varid{args}\;(\Varid{map}\;(\Varid{patS}\;\Varid{whnf})\;\Varid{ps}){}\<[E]%
\ColumnHook
\end{pboxed}
\)\par\noindent\endgroup\resethooks
\end{spaced}%
Function \ensuremath{\Varid{genEvalStrat}} takes a rule identifier, a function definition and
the weak-head normal form strategy and returns an evaluation strategy for 
that function definition. First, arguments are evaluated by \ensuremath{\Varid{patListS}}. 
Second, the generated rewrite rule for the definition is applied.
\begin{spaced}\begingroup\par\noindent\advance\leftskip\mathindent\(
\begin{pboxed}\SaveRestoreHook
\column{B}{@{}>{\hspre}l<{\hspost}@{}}%
\column{37}{@{}>{\hspre}c<{\hspost}@{}}%
\column{37E}{@{}l@{}}%
\column{42}{@{}>{\hspre}l<{\hspost}@{}}%
\column{E}{@{}>{\hspre}l<{\hspost}@{}}%
\>[B]{}\Varid{genEvalStrat}\mathbin{::}\Conid{Id}\to \Conid{Def}\to \Conid{Strategy}\;(\Conid{Context}\;\Conid{Expr})\to \Conid{Strategy}\;(\Conid{Context}\;\Conid{Expr}){}\<[E]%
\\
\>[B]{}\Varid{genEvalStrat}\;\Varid{rId}\;(\Conid{Def}\;\Varid{s}\;\Varid{ps}\;\Varid{e})\;\Varid{whnf}{}\<[37]%
\>[37]{}\mathrel{=}{}\<[37E]%
\>[42]{}\Varid{patListS}\;\Varid{whnf}\;\Varid{s}\;\Varid{ps}{}\<[E]%
\\
\>[37]{}\mathrel{{<}\hspace{-0.4em}\star\hspace{-0.4em}{>}}{}\<[37E]%
\>[42]{}\Varid{liftToContext}\;(\Varid{genRule}\;\Varid{rId}\;(\Conid{Def}\;\Varid{s}\;\Varid{ps}\;\Varid{e})){}\<[E]%
\ColumnHook
\end{pboxed}
\)\par\noindent\endgroup\resethooks
\end{spaced}%
The generated evaluation strategy defined above does not behave correctly for 
multiple or nested patterns. The strategy \ensuremath{\Varid{s}\mathrel{{<}\hspace{-0.4em}\star\hspace{-0.4em}{>}}\Varid{t}} only succeeds if both
\ensuremath{\Varid{s}} and \ensuremath{\Varid{t}} succeed. In the context of an evaluation strategy this is 
undesirable because the effect of partly evaluating expressions with strategy 
\ensuremath{\Varid{s}} is undone (and cannot be observed) if strategy \ensuremath{\Varid{t}} fails. The solution is 
to replace all occurrences of \ensuremath{\mathrel{{<}\hspace{-0.4em}\star\hspace{-0.4em}{>}}} (and derived combinators such as \ensuremath{\Varid{sequence}}) 
in the generated strategy by a new combinator \ensuremath{\mathrel{{<}\hspace{-0.2em}\star}}, which we define as:
\begin{spaced}\begingroup\par\noindent\advance\leftskip\mathindent\(
\begin{pboxed}\SaveRestoreHook
\column{B}{@{}>{\hspre}l<{\hspost}@{}}%
\column{E}{@{}>{\hspre}l<{\hspost}@{}}%
\>[B]{}(\mathrel{{<}\hspace{-0.2em}\star})\mathbin{::}\Conid{Strategy}\;\Varid{a}\to \Conid{Strategy}\;\Varid{a}\to \Conid{Strategy}{}\<[E]%
\\
\>[B]{}\Varid{s}\mathrel{{<}\hspace{-0.2em}\star}\Varid{t}\mathrel{=}\Varid{s}\mathrel{{<}\hspace{-0.4em}\star\hspace{-0.4em}{>}}(\Varid{t}\mathrel{\triangleright}\Varid{succeed}){}\<[E]%
\ColumnHook
\end{pboxed}
\)\par\noindent\endgroup\resethooks
\end{spaced}%

\section{Prototype}
\label{section:feedback}

\begin{figure}[t]
\hfill\includegraphics[height=6.5cm]{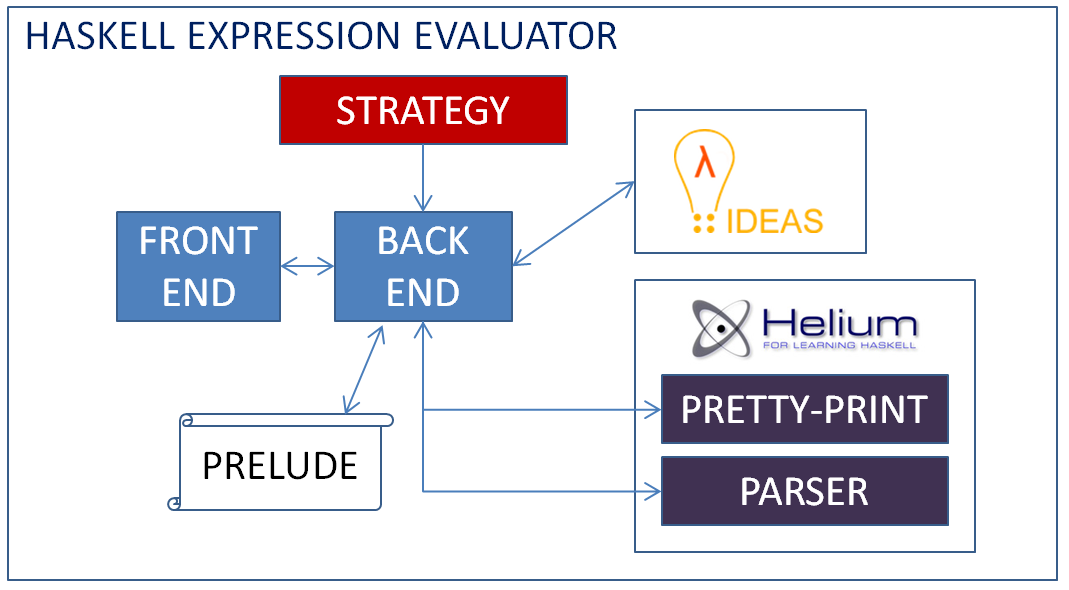}\hfill\mbox{}
\caption{Component diagram of the prototype}
\label{fig:architecture}
\end{figure}

The prototype for practicing with evaluation steps is divided into a number of separate components.
Figure~\ref{fig:architecture} shows the component model of the prototype: it consists of a front-end, a back-end, and
a strategy component. 
The strategy component contains all rewrite rules and rewrite strategies for a
certain evaluation strategy. 
The back-end component uses the external components \ideas{}
and Helium, and reads the prelude file with function definitions that resides 
on the server. This paper focuses on the back-end and strategy components.

In the future we hope to integrate the evaluator with the \askelle{}
programming tutor~\cite{gerdes2012ask} for learning Haskell. This tutor supports students in
solving introductory programming exercises by providing hints on how to 
continue, and by diagnosing intermediate programs that are submitted by a student. 
Having a stepwise evaluator in the \askelle{} environment is promising 
because it allows students to see their programs being evaluated, and to 
pinpoint mistakes in their definitions (e.g.\ by evaluating a counter-example).
\askelle{}
is also build on top of the \ideas{} framework and the Helium compiler.
Decomposing the evaluator into components, some of which are shared with
\askelle{}, should make future integration easier.

\paragraph{Front-end.}

The front-end is web-based and written in HTML and JavaScript. It uses JSON to communicate with the back-end. It provides an interface
to inspect the evaluation of a Haskell expression or to practice with the
evaluation of a Haskell expression. The prototype front-end can easily be
replaced by another front-end and the purpose of this prototype front-end is to
show what kind of \ideas{} services~\cite{Heeren2014110} can be used. 

A user can
select an example Haskell expression by clicking on the `Select' button (see 
Figure~\ref{fig:prototype}) or she can
enter a Haskell expression. The prototype currently only supports a
subset of the Haskell syntax, so it is possible that this operation fails. After
typing or selecting a Haskell expression the user can choose between the
innermost evaluation strategy and the outermost evaluation strategy. The user can
now call several standard \ideas{} services such as the service for calculating
the number of steps left, getting information about the next rule that should be applied, or
finding out what the result is of applying the rule. The user can fill in the next
evaluation step, possibly with the help of the services, and click on the
button `Diagnose'  to see if the provided next step is the correct step
according to the strategy.

The string representation of a rule is used by the feedback service 
that gives information about the next rule that should be applied. This string representation
can be modified in a script file~\cite{Heeren2014110} where rule identifiers are mapped to
textual representations. Every rewrite rule has an identifier. For
example, the identifier of the \ensuremath{\Varid{foldl}} rewrite rule is \ensuremath{\text{\tt \char34 eval.foldl.rule\char34}} (Section~\ref{section:rewriterules}). 
This identifier is mapped to `Apply the fold left rule to process a
list using an operator that associates to the left' in the script file. This script file can be
changed without recompiling the evaluator, which makes it possible to easily
adapt the information, for example to support feedback in another language.

\paragraph{Back-end.}
The back-end is developed in Haskell, and uses the Helium compiler~\cite{Helium} 
for parsing and pretty-printing expressions. 
The back-end operates as a glue component that connects all
other components. It receives a string from the front-end, uses the
Helium compiler to parse the string, and converts the abstract syntax tree produced by Helium to
the expression datatype. The back-end receives expression results from
\ideas{}, converts values from the expression datatype back to Helium's expression
datatype, and uses the Helium compiler to convert this value to a
string. This string is presented to the user using the front-end.

To determine if a provided step follows
the evaluation strategy, the \ideas{} framework is instantiated with functionality for determining if the provided
expression is equal to the expected expression. The evaluator therefore needs to
implement two functions that are required by the framework: one function 
to determine if two expressions are
semantically equivalent, and one function to determine if two expressions are
syntactically equivalent. 
Syntactic equivalence is obtained by deriving an 
instance of the \ensuremath{\Conid{Eq}} type class for the \ensuremath{\Conid{Expr}} datatype. 
Semantic equivalence is more subtle because a student may submit an
expression that is syntactically different from the expression we expect to get from the step, but semantically the 
same. Semantic equivalence is defined by using a function that applies the rewrite strategy until it cannot be 
applied any further and returns \ensuremath{\Conid{True}} if both results are the same.
With these two functions we can also spot mistakes in rewrite steps, even when an
incorrect expression happens to produce the same result.

\paragraph{Survey.}

We carried out a small survey to find out if the tool has potential in an 
educational setting. We invited instructors of functional programming courses at several 
universities, primarily but not only in the Netherlands, and students that follow or completed the functional programming
course at the Open Universiteit to experiment with the prototype,
and to answer open-ended and closed-ended questions about the tool. All participants were 
approached via email on April~30, 2014. By May~12, 7 instructors (out of 8) and
9 students (out of 29) completed the survey. All questions and answers are
included in the first author's master's thesis~\cite{OlmerThesis}. We
discuss the most important observations.

All participants agree that it is useful for students following an introductory 
course in functional programming to inspect how an expression is evaluated. 
Eleven participants agree that it is also useful to inspect multiple evaluation strategies. 
One student argued that the examples in Hutton's textbook~\cite{Hutton} are 
sufficient to get a good understanding of the evaluation steps, and two participants 
argued that inspecting evaluation steps according to multiple evaluation strategies 
may be confusing for those who start programming for the first time. 
There is less consensus about the question if more evaluation strategies should be supported. 
Four instructors argue that lazy evaluation should be supported because 
it is very subtle, but two other instructors argue that this is too 
much for the basic understanding of programming concepts. Another instructor thinks it would be nice to add lazy evaluation, but is not sure if this is worth the effort. 
Students find the addition of lazy evaluation useful. 

Practicing with the evaluation 
steps was well received by all participants. Three participants argue that it is only 
useful at the very start, and that students can quickly become bored by entering all evaluation 
steps. Another remark from an instructor is that studying evaluation steps is 
not the only way to reason about a program. To ensure that students do not get bored quickly 
it might be useful to skip certain evaluation steps. However, participants disagree about whether or not this feature should be supported. Five instructors
and one student argue that it should not be possible to skip steps to illustrate that each step is necessary. The other participants would like to see 
this feature added to the prototype. 

Nine participants find it possibly useful to add user-defined function definitions and 
to adjust function definitions to be able to inspect the effect of these changes on the evaluation of 
expressions. Three participants do not find this useful and think support for prelude functions is sufficient. 
Some participants suggest to also show the function 
definitions. This could help students in understanding the evaluation steps. Another suggestion 
is to show a derivation according to multiple evaluation strategies side by side, so that the student 
can easily observe the differences. Some participants also suggest to detect if an expression 
cannot be evaluated according to the innermost strategy and notify the user accordingly.

It is clear that instructors not always agree upon desirability of features. 
Therefore, it is important that features are configurable for instructors, and that an instructor 
can adapt the tool to her own course.

\section{Related work}
\label{section:relatedwork}

There are roughly three approaches to inspect the evaluation steps of a Haskell
expression: trace generation, observing intermediate data structures, and using
rewrite rules. The central idea of the trace generation approach, which is
mainly used for debugging, is that every expression is transformed into an
expression that is supplemented with a description in the trace. The trace
information is saved in a datatype that can be viewed by a trace viewing
component. There are two methods for trace generation. The first method is to
instrument Haskell source code. Pure Haskell functions are transformed to
Haskell functions that store the evaluation order in a certain datatype that
can be printed to the user. This approach is used to show complete traces in
so called redex trail format~\cite{sparud:redex}, and this is also the approach
used in the Hat debug library~\cite{chitil:hat}. An advantage of this method is
that it is completely separated from the compiler, so it does not matter which
compiler is used. A disadvantage of this method is that the instrumentation of
the original code can alter the execution of the program. The second method,
which is used for example by WinHIPE~\cite{Pareja-Flores:winhipe}, is to
instrument the interpreter. The advantage of this method is that the execution
of the program is exactly the same, but a disadvantage is that the interpreter
(part of the compiler) needs to be adjusted. The approach to observe
intermediate data structures, which is also mainly used for debugging, is used in
the Hood debugger~\cite{sparud:redex}. The approach to specify rewrite rules to
inspect the evaluation of expressions is used for example in the stepeval
project where a subset of Haskell expressions can be inspected~\cite{StepEval}.
With the above approaches it is possible to inspect the evaluation steps of a
Haskell expression, but it is not possible to practice with these evaluation
steps.

Several intelligent tutoring systems have been developed that support students with
learning a functional programming language. One of the main problems for novice
programmers is to apply programming concepts in practice~\cite{Lahtinen:novice}.
To keep students motivated to learn programming it is therefore important to
teach it incrementally, to practice with practical exercises, and to give them
early and direct feedback on their work~\cite{Vihavainen}. 
The main advantage of an intelligent
tutoring system is that a student can get feedback at any moment. An
intelligent tutoring system consists of an inner loop and an outer loop. The
main responsibility of the outer loop is to select an appropriate task for the
student; the main responsibility of the inner loop is to give hints and
feedback on student steps. The Web-Based Haskell Adaptive Tutor (WHAT) focuses more on
the outer loop. It classifies each student into a group of students that share
some attributes and will behave differently based on the group of the
student~\cite{WHAT}. With WHAT, a student can practice with three kinds of problems:
evaluating expressions, typing functions, and solving programming assignments. A
disadvantage of this tutor is that it does not support the stepwise development
of a program. \askelle{}~\cite{gerdes2012ask} is a Haskell tutor system that focuses primarily on the
inner loop. Its goal is to help students learn functional programming by developing
programs incrementally. Students receive feedback about whether or not they are on the
right track, can ask for a hint when they are stuck, and see how a complete
program is constructed stepwise~\cite{cefp}.

\section{Conclusions and future work}
\label{section:conclusion}

In this paper we have presented a prototype tool that can be used to show the
evaluation steps of a Haskell expression according to different evaluation
strategies. We support the left-most innermost evaluation strategy and an 
outermost evaluation strategy based on pattern matching.
The tool can also be used to practice with evaluation steps.
This prototype may help students to better understand important programming
concepts such as pattern matching, higher-order functions, and recursion. How 
effective the tool is in practice needs to be further investigated 
by collecting more empirical data.

The evaluation process is driven by the definition of the rewrite rules and the
evaluation strategy that is used. 
The rewrite rules and evaluation strategies for
the definitions can be generated by the prototype from a set of function definitions. 
The extension of adding user-defined function definitions makes it possible
for users to easily get an understanding of how their function will be
evaluated. Another advantage is that alternative definitions for prelude functions 
can be tried, and that the results can be inspected. For example, a student can 
use a strict version of \ensuremath{\Varid{foldl}} for \ensuremath{\Varid{sum}}, or define \ensuremath{\Varid{sum}} recursively.
At the moment, adding user-defined functions requires an instructor to change the
content of the prelude file on the server. Preferably, some support for this
feature is added to the front-end.

The expression language in the current tool is limited. In the 
future, we hope to support other language constructs such as guards, 
conditional expressions, list comprehensions, algebraic datatypes, etc.
Other future work is to offer the possibility to change the step size of a function,
and to add sharing to an evaluation strategy:
\begin{itemize}
\item
In the examples, we use a fixed step size of one. The step size is 
the number of steps that the evaluator uses to rewrite a certain expression. For
example, the expression \ensuremath{\mathrm{3}\mathbin{+}(\mathrm{4}\mathbin{+}\mathrm{7})} is evaluated to \ensuremath{\mathrm{14}} in two steps, although 
most students will typically combine these steps. More research must be carried out
to automatically derive or configure a certain step size that suits most students.
\item
The lazy evaluation strategy used by Haskell combines the outermost 
evaluation strategy with sharing. Currently, sharing is not supported in the 
prototype. To help students learn about which computations are shared, we plan 
to extend the prototype along the lines of Launchbury's natural semantics for lazy evaluation~\cite{Launchbury93anatural}, and by making the heap explicit.
\end{itemize}

The long-term goal of our work is to integrate the functionality of the
prototype in the \askelle{} programming tutor, which then results in a complete
tutoring platform to help students learn programming. 
We are also considering to combine the 
evaluator with QuickCheck properties~\cite{QuickCheck}: when QuickCheck 
finds a minimal counter-example that falsifies a function definition (e.g.\ for
a simple programming exercise), then we can use the evaluator to explain more
precisely why the result was not as expected, or use the evaluator as a
debugging tool.

\paragraph{Acknowledgements}
The authors would like to thank the students and instructors that participated 
in the survey for their feedback and suggestions. We also thank the anonymous 
reviewers for their detailed comments.

\bibliographystyle{eptcs}
\bibliography{eval}

\end{document}